\documentclass[
journal]{IEEEtran}  

\newcommand{\bm}[1]{\mbox{\boldmath{$#1$}}}
\usepackage{graphicx}
\graphicspath{undefined{figures/},} 
\usepackage{stfloats}
\usepackage{graphicx}
\usepackage{epstopdf}
\usepackage{bbm}
\usepackage{amsfonts}
\usepackage{amsmath}
\usepackage{CJK}
\usepackage{thmtools}
\usepackage{mathrsfs}
\usepackage{multirow}
\usepackage{booktabs}
\usepackage{setspace}
\usepackage{enumerate}
\usepackage{amssymb,amscd,amsmath}
\usepackage[all]{xy}
\usepackage{fancyhdr}
\usepackage{subfigure}
\usepackage{bm}
\usepackage{algorithm}
\usepackage{algorithmic}
\usepackage{caption}
\usepackage{stfloats}
\usepackage[utf8]{inputenc}
\usepackage[T1]{fontenc}
\usepackage{url}
\usepackage{ifthen}
\usepackage{cite}
\usepackage{thmtools}
\usepackage{amssymb,mathrsfs}
\usepackage{amsmath}
\usepackage{amsthm}
\usepackage{amstext}
\theoremstyle{} 
\newtheorem{theorem}{Theorem} [section]

\newtheorem{remark} [theorem]{Remark}

\allowdisplaybreaks[4]
\usepackage{xcolor}
\usepackage{array}
\usepackage[caption=false,font=normalsize,labelfont=sf,textfont=sf]{subfig}
\usepackage{textcomp}
\usepackage{verbatim}
\def\BibTeX{{\rm B\kern-.05em{\sc i\kern-.025em b}\kern-.08em
    T\kern-.1667em\lower.7ex\hbox{E}\kern-.125emX}}

\begin{document}
\title{Random Sampling of Bandlimited Graph Signals from Local Measurements}
\author{Lili Shen, Jun Xian, Cheng Cheng
\thanks{Shen and Cheng are with the School of Mathematics, Sun Yat-sen University, Guangzhou 510275, China (e-mail: shenlli@mail2.sysu.edu.cn; chengch66@mail.sysu.edu.cn). Xian is with the School of Mathematics and Guangdong Province Key Laboratory of Computational Science, Sun Yat-sen University, Guangzhou 510275, China (e-mail: xianjun@mail.sysu.edu.cn). This project is partially supported by the National Natural Science Foundation of China (11631015, 11871481, 12026601, 12171490); Guangdong Province Nature Science Foundation (2022A1515011060); Guangzhou Science and Technology Foundation Committee (202201011126); the Guangdong Provincial Government of China through the Computational Science Innovative Research Team program, China; the Guangdong Province Key Laboratory of Computational Science, China.}
}
\maketitle
\begin{abstract}
The random sampling on graph signals is one of the fundamental topics in graph signal processing. In this letter, we consider the random sampling of $k$-bandlimited signals from the local measurements and show that no more than $O(k\log k)$ measurements with replacement are sufficient for the accurate and stable recovery of any $k$-bandlimited graph signals. We  propose two random sampling strategies based on the minimum measurements, i.e., the optimal sampling and the estimated sampling. The geodesic distance between vertices is introduced to design the sampling probability distribution. Numerical experiments are included to show the effectiveness of the proposed methods.
\end{abstract}


\IEEEpeerreviewmaketitle
 \vspace{-1em} 
\section{Introduction}\label{introduction}
Graph provides an innovative tool to represent the complicated networks, such as social networks, wireless sensor networks, and brain networks \cite{wire,chengc}. The emerging field of graph signal processing (GSP) addresses the limitations that many tools in the classical signal processing handling the signals residing in the regular domain can not be easily generalized to the irregular domain, such as the  sampling theorem, Fourier transform, wavelet, etc. \cite{shannon,fourier1,wave1,gsp1,gsp2,gsp3,R1,R2,R3}.  


Sampling and reconstruction of bandlimited graph  signals is one of the most popular topics in GSP \cite{R1,R2,Pes,AA}. 
Deterministic sampling for bandlimited graph signals has been widely studied, which depends on a delicate selection of a fixed node subset to ensure the stable reconstruction \cite{AA,Pes,sihengchen,Yang,Diego,ls2015,ls2016,lsYang,Jiang1,ave,clusters}. The authors in \cite{Pes} introduced the notion of uniqueness set with vertex-wise sampled data. Following the work of \cite{Pes}, the authors in \cite{AA,sihengchen,Yang} showed that a uniqueness set of size $k$ that enables perfect reconstruction of $k$-bandlimited signals always exists. However, it is computationally expensive to find such the deterministic sampling set \cite{R1,NPhard}. To address this challenge, the idea of random sampling in which sampled vertices are drawn by a certain probability distribution was considered in \cite{sihengchen,gill1,gill2,RV,Tre}. In particular,  
the authors in \cite{sihengchen} proved that the uniform  sampling is optimal for the Erd{\"o}s-R\'{e}nyi graph.  
 Later, the authors in \cite{gill1,gill2}  showed that the sampling probability distribution based on the graph structure performs better than the uniform distribution in reducing the number of measurements. 
This designed random sampling strategy was further extended to signals residing on product graphs \cite{RV}.  


 In practical applications, the sampled data may not be vertex-wise and dependent on a linear combination of signals on local sets due to the physical obstacles \cite{chengc,Jiang,cheng1,JiangTV,JiangD}. Sampling from local weighted measurements has many applications in the wireless sensor networks with clustering features, such as the environment monitoring in which the samples are linear combinations of the signal over a local set of vertices \cite{WSN1,WSN2}. The authors in \cite{ls2016} studied the deterministic sampling and reconstruction from local weighted measurements. Further researches on sampling and reconstruction of graph signals from samples collected on local sets can be referred to \cite{ls2015,ls2016,Yang,R1,Diego,clusters}.  
 
In this letter, we consider the random sampling of bandlimited graph signals from local measurements. The letter is organized as follows. In Section \ref{sec3}, a random sampling scheme based on the  local measurements is proposed, and a sufficient condition is provided to ensure the recovery of bandlimited graph signals with high probability, see Theorem \ref{main}. In Section \ref{sec4}, an optimal probability distribution and its estimator are presented. By considering the geodesic distances between vertices, a reordered sampling distribution is further introduced to effectively reduce the redundancy of local measurements, see Algorithm \ref{code:recentEnd}.  In Section \ref{sec6}, simulations are conducted to verify the effectiveness of the proposed random sampling scheme from local measurements. In Section \ref{sec7}, we conclude this letter.
 \vspace{-0.5em} 
\section{Random Sampling Method with Locally Weighted Measurements} \label{sec3} 
In this section, we first introduce some  preliminaries on the $k$-bandlimted graph signals,  and then present a locally weighted random sampling procedure.  
We show that the proposed locally weighted random sampling procedure can stably embed the set of $k$-bandlimited graph signals. 
 \vspace{-1em} 
\subsection{Preliminaries}
 Denote the matrices, vectors, sets and scalars by bold capital letters, bold lowercase letters, calligraphic uppercase letters, and regular letters, respectively. Let $|\cdot|$ denote the cardinality of a set, and $\lceil \cdot\rceil$  denote the ceiling function of some scalars. 
Let $\mathcal{G} =({\mathcal V, \mathcal{E}}, {\bf W})$ denote a weighted undirected simple graph with vertex set $\mathcal{V} =\{1,2,\dots,n\}$, edge set $\mathcal{E} \subset \mathcal{V} \times \mathcal{V}$ and weighted adjacency matrix ${\bf W} \in \mathbb{R}^{n\times n}$. The \textit{geodesic distance} $d(i,j)$ is the number of edges in the shortest path connecting vertices $i$ and $j$. 
The \textit{Laplacian matrix} of $\mathcal{G}$ is defined as ${\bf L} ={\bf D} -{\bf W},$ where ${\bf D}$ is a diagonal matrix with element $d_{i} =\sum_{j\in \mathcal{V}} w_{ij}, i\in \cal V$, and ${\bf W} =(w_{ij})_{1\leqslant i, j\leqslant n}$ is a symmetric matrix with element $w_{ij} >0$ if $i$ and $j$ are adjacent and $0$ otherwise. Write the eigendecomposition of the Laplacian  ${\bf L}={\bf U}{\bf\Lambda}{\bf U}^{\top}$, where ${\bf\Lambda}$ is the diagonal matrix with eigenvalues  of $\bf L$ deployed on the diagonal in ascending order
 and ${\bf U}$ is an orthogonal matrix with column vectors being the corresponding orthonormal eigenvectors ${\bm u}_i$.
The graph signal ${\bf x}$ is represented by a vector that is indexed on the vertices of the graph $\mathcal{G}$, and the \textit{$k$-bandlimited graph signal} is defined as ${\bf x} \in \text{span}({\bf U}_k)$, where ${\bf U}_k=({\bf u}_1,{\bf u}_2,\dots,{\bf u}_k)$ contains the first $k$ columns of matrix ${\bf U}$. Throughout this letter, it is assumed that the bandwidth $k$ is given, and $\lambda_k\neq\lambda_{k+1}$.
\vspace{-1em} 
 \subsection{Random Sampling Scheme} \label{schemesec} 
Let each vertex $i\in {\mathcal V}$ be selected with probability $p_i$, where $p_i>0$ and $\sum_{i=1}^n{p_i} =1$. The sampling set $\Omega=\{{\omega_1},{\omega_2},\dots,{\omega_m} \}$ is constructed by drawing $m$ nodes independently  (with replacement) from a vertex set with probability distribution ${\bf p} =(p_1, \dots, p_n)^{T}$,  where $m$ is  the number of measurements.
Let ${\bf \Phi}=({\pmb \varphi}_1, {\pmb \varphi}_2,\dots,{\pmb \varphi}_n)^\top$ denote a locally weighted matrix that is obtained by a polynomial of the Laplacian matrix ${\bf L}$, i.e.,
\begin{equation} \label{phi2} 
\begin{aligned} 
{\bf \Phi}:&=g({\bf L})=\sum_{i=0}^L\alpha_i{\bf L}^i={\bf U} g({\bf \Lambda}){\bf U}^{T},
 \end{aligned} 
\end{equation} 
where $g(t)=\sum_{i=0}^L\alpha_i{t}^i$ is a univariate polynomial and the coefficients $\alpha_i$ is selected with $g(t)\neq 0$ for all $t\in \bf \Lambda$. 
The sampled data ${\bf y}$ is obtained by  
${\bf y} ={\bf\Psi}{\bf x}\in {\mathbb R}^m$,
where the locally weighted sampling matrix 
\begin{equation} \label{lwsm} 
\begin{aligned} 
 {\bf\Psi} :
=({\pmb \varphi}_{\omega_1}, {\pmb \varphi}_{\omega_2},\dots,{\pmb \varphi}_{\omega_m})^\top
 \end{aligned} 
\end{equation} 
 is a sub-matrix of  $\bf \Phi$ in \eqref{phi2} with rows indexed by the sampling set  $\Omega$. Since the geodesic width of ${\bf \Phi}$ is no more than the degree $L$ of the polynomial, the observation at a sampled node $\omega_i\in \Omega$  is a linear combination  within $L$-neighborhood, i.e., 
\begin{equation} \label{sampleddata} 
\begin{aligned} 
{y}_i :=\langle{\bf x},{\pmb \varphi}_{\omega_i} \rangle={\pmb \varphi}_{\omega_i}^{T} {\bf x} =\sum_{j\in \overline{\mathcal{N}}_{\omega_i}} \varphi_{\omega_i j} x_j,
 \end{aligned} 
\end{equation} 
where the local set $\overline{\mathcal{N}}_{\omega_i}$ is composed by $\{j |d(\omega_i,j)\leqslant L\} $. More merits of matrices of finite geodesic width can be referred to \cite{Jiang,cheng1}.

%

The quantity $\|{\bf U}_k^{T}{\pmb \varphi}_i\|_2^2$ is used to characterize the energy of the local set $\overline{\mathcal{N}}_{i}$ concentrated on the first $k$ Fourier modes.  It is natural to have that the large (or small) sampling distribution $p_{\omega_i}$ is related to the large (or small) $\|{\bf U}_k^{T}{\pmb \varphi}_{\omega_i}\|_2^2$ at the sampling node $\omega_i$ \cite{gill1}.  
In the following theorem, we show that a $k$-bandlimited graph signal can be recovered with high probability, provided that the number of local measurements is sufficient large.
Due to the space limit, its detailed proof can be found in the Supplementary Material.
\begin{theorem} \label{main} 
Let $\Omega=\{{\omega_1},{\omega_2},\dots,{\omega_m} \}$ be the sampling set drawing independently with replacement by probability distribution $\bf p$, and the sampling distribution matrix ${\bf P}_\Omega=\text{diag} (p_{\omega_1},p_{\omega_2},\dots,p_{\omega_m})$. Let the locally weighted sampling matrix ${\bf\Psi}$ be in \eqref{lwsm},  $c_1=\min_{1\leqslant i\leqslant k} \{(g(\lambda_i))^2\}$ and $c_2=\max_{1\leqslant i\leqslant k} \{(g(\lambda_i))^2\}$. Then, for any $\varepsilon, \delta\in(0,1)$, 
\begin{equation} \label{frame1} 
\begin{aligned} 
(1-\delta)c_1\|{\bf x} \|_2^2\leqslant \frac{1}{m} \|{\bf P}_\Omega^{-\frac{1}{2}}{\bf\Psi}{\bf x} \|_2^2\leqslant(1+\delta)c_2\|{\bf x} \|_2^2\\
\end{aligned} 
\end{equation} 
holds with probability at least $1-\varepsilon$ for all ${\bf x} \in {\rm span}({\bf U}_k)$ provided that the number of measurements
\begin{equation} \label{m} 
\begin{aligned} 
m\geqslant \frac{3 \zeta_\Omega}{c_1\delta^2} \log\frac{2k}{\varepsilon},
\end{aligned} 
\end{equation} 
where $\zeta_\Omega=\max_{\omega_i\in \Omega} \{{\frac{\|{\bf U}_k^{T}{\pmb \varphi}_{\omega_i}\|^2_2}{p_{\omega_i}}}\}$.
\end{theorem}
Note that the graph weighted coherence 
\begin{equation}\label{Gcoherence.def}
\begin{aligned} 
\zeta_\Omega=\max_{\omega_i\in \Omega} \{{\frac{\|{\bf U}_k^{T}{\pmb \varphi}_{\omega_i}\|^2_2}{p_{\omega_i}}}\}
\end{aligned} 
\end{equation} 
 represents how the energy of these signals spreads over the sampled nodes, and it is essential for the number of local measurements which enables the stable sampling of $k$-bandlimited graph signals. 
%
\begin{remark} \label{oklogk}{\rm 
By the definition of $\zeta_\Omega$ in \eqref{Gcoherence.def} and the randomness of $\Omega$, if the locally weighted matrix $\bf \Phi$ is as in \eqref{phi2}, the number of measurements in \eqref{m} will be
$m\geqslant \frac{3 \zeta }{c_1\delta^2} \log\frac{2k}{\varepsilon} $,
where 
\begin{equation} \label{zeta}
\begin{aligned} 
\zeta&=\hskip-.1in&\hskip-.1in\max\limits_{i\in \mathcal{V}} \Big\{\frac{\|{\bf U}_k^{T}{\pmb \varphi}_{i} \|_2^2}{p_i} \Big\}=\max\limits_{1\leqslant i \leqslant n} \{\frac{\|{\bf U}_k^{T}{\pmb \varphi}_i\|_2^2}{p_i} \} \sum_{i=1}^n p_i \\
&\geqslant&\sum_{i=1}^n p_i\frac{\|{\bf U}_k^{T}{\pmb \varphi}_{i} \|^2}{p_i}=\sum_{i=1}^n\|{\bf U}_k^{T}{\pmb \varphi}_i\|^2_2=\sum_{j=1}^k(g(\lambda_j))^2.
\end{aligned} 
\end{equation} 
Thus, we have
\begin{equation} \nonumber
\begin{aligned} 
m\geqslant \frac{3 \zeta }{c_1\delta^2} \log\frac{2k}{\varepsilon} &\geqslant \frac{3 \sum_{j=1}^k(g(\lambda_j))^2 }{\min\limits_{1\leqslant j\leqslant k} \{(g(\lambda_j))^2\} \delta^2} \log\frac{2k}{\varepsilon} \geqslant \frac{3 k }{\delta^2} \log\frac{2k}{\varepsilon}, 
\end{aligned} 
\end{equation} 
which implies that $O(k\log k)$ measurements are sufficient for \eqref{frame1}. 
} 
\end{remark} 
 \vspace{-1em} 
\section{Sampling Probability Distributions} \label{sec4}
In this section, we first study the optimal probability distribution in \eqref{optiP} and the estimated probability distribution in \eqref{estP}. We later propose a  new sampling probability distribution in Algorithm \ref{reA} based on the geodesic distance and the local set energy, which reduces the overlapping of local measurements caused by the randomness of samples.   
\vspace{-1em}
\subsection{Optimal Sampling Distribution} 
Theorem \ref{main} provides a sufficient condition for the reconstruction of $k$-bandlimited graph signals from sufficiently large measurements that are taken on the nodes obeying an arbitrary sampling probability distribution. In order to minimizing the measurements number, a proper sampling distribution is designed to reach the lower bound of $\zeta_\Omega$  in \eqref{Gcoherence.def}. Based on the locally weighted matrix ${\bm \Phi}$ in \eqref{phi2}, an optimal sampling distribution ${\bf p}_{\rm opt}=( p_1^*,\dots, p_n^*)$ with element
  
 \begin{equation} \label{optiP} 
\begin{aligned} 
 p_i^*=\frac{\|{\bf U}_k^{T}{\pmb \varphi}_i\|_2^2}{\sum_{j=1}^k(g(\lambda_j))^2} \;\;\text{for $i=1,2,\dots, n$}
 \end{aligned} 
\end{equation} 
 can be obtained by the fact that the equality in \eqref{zeta} holds if and only if
 $\max\limits_{i\in \mathcal{V}} \Big\{\frac{\|{\bf U}_k^{T}{\pmb \varphi}_i\|_2^2}{p_i} \Big\} =\min\limits_{i\in\mathcal{V}}\Big \{\frac{\|{\bf U}_k^{T}{\pmb \varphi}_i\|_2^2}{p_i} \Big\}.$
\vspace{-1em} 
\subsection{Estimated Sampling Distribution} \label{est.propose}
The optimal sampling probability distribution in \eqref{optiP} needs to know the eigenvectors corresponding to the first $k$ smallest eigenvalues of the Laplacian matrix, which is computationally expensive and may not work for the graph of large size. 
The following theorem can be adapted from \cite{gill1} by using Gaussian random variables to estimate the optimal sampling probability distribution. 
\begin{theorem} \label{estUk} 
Let ${\bf r}^1,{\bf r}^2,\dots,{\bf r}^t\in\mathbb{R}^n$ be $t$ independent zero-mean Gaussian random vectors with covariance $\frac{1}{t}{\bf I}$. 
Then, there exists an absolute constant $\gamma>0$ such that for any $\varepsilon, \delta\in(0,1)$, with probability at least $1-\varepsilon$, 
\begin{equation} \label{Ukineq} 
(1-\delta)\|{\bf U}_{k}^T{\pmb \varphi}_i\|_2^2\leqslant \sum_{\ell=1}^t \langle {\bf r}^\ell_{b_{\lambda_k}},{\pmb \varphi}_i\rangle^2 \leqslant(1+\delta)\|{\bf U}_{k}^T{\pmb \varphi}_i\|_2^2
\end{equation} 
for all $i\in\{1,2,\dots,n\}$ provided that
$t\geqslant\frac{\gamma}{\delta^2} \log\frac{2n}{\varepsilon}$, where ${\bf r}^\ell_{b_{\lambda_k}}:={\bf U}b_{{\lambda_k}}(\text{diag}({\bm\Lambda})){\bf U}^T{\bf r}^\ell$ denotes the Gaussian random variable ${\bf r}^\ell$ after filtering by the ideal low-pass filter ${b_{{\lambda_k}} }$ with cut-off frequency ${\lambda_k} >0$.  
\end{theorem} 
Note that ${\bf r}^\ell_{b_{\lambda_k}}$ can be approximated by the Chebyshev polynomial expansion instead of Laplace eigendecomposition. The estimated distribution ${\bf p}_{\rm est}=(\overline{p}_1,\dots,\overline{p}_n)$ can be obtained with
\begin{equation} \label{estP} 
\begin{aligned} 
\overline{p}_i=\frac{\sum_{\ell=1}^t{\langle {\bf r}^\ell_{b_{\lambda_k}}, {\pmb \varphi}_i\rangle^2}}{\sum_{i=1}^n\sum_{\ell=1}^t{\langle {\bf r}^\ell_{b_{\lambda_k}}, {\pmb \varphi}_i\rangle^2}}.
\end{aligned} 
\end{equation} 
This implies that $O(n\log{n})$ complexity is sufficient to realize the estimated distribution during the sampling preparation.
\vspace{-1em} 
 \subsection{Reordered Sampling Distribution}
Since both \eqref{optiP}  and its estimator \eqref{estP} build upon the energy of the local sets, the intersection of the local sets may lead to the redundancy of local measurements. 
In order to reduce the redundancy of local measurements,  we propose a reordered sampling probability distribution in Algorithm \ref{reA}, in which we employ the distance coherence algorithm \cite{dist} to  incorporate vertex distances into the sampling probability distribution. 

Define the distance between vertex set $\mathcal{V}_1$ and vertex $i$ by  $d(\mathcal{V}_1,i)=\min \{d(u,i) |u\in \mathcal{V}_1\}$, and let $\mathcal{D}_{\mathcal{V}_1}$ contain the vertices  $i$ satisfying $d(\mathcal{V}_1,i)>L$. 
 \begin{algorithm} [h]
\caption{The Reordered Sampling Distribution Based on Distance} \label{reA} 
\textbf{Input:}{ Graph $\mathcal{G}$; the original probability distribution ${\bf q}=(q_1, q_2,\dots, q_n)$ (${\bf q}={\bf p}_{\rm opt}$ or ${\bf p}_{\rm est}$); the positive integer $L$}.\\
\textbf{Initialization:}{ Reorder the vertices $\alpha_i$ by the value of $q_i$ such that ${\bf q}_1^\ast=(q_{\alpha_1}, q_{\alpha_2},\dots, q_{\alpha_n})$ where $q_{\alpha_i} \geqslant q_{\alpha_j}$ for all $i\leqslant j$; 
Set $p_{\alpha_1} =q_{\alpha_1}$;
$\mathcal{V}_1=\{\alpha_1\}$; 
$\mathcal{D}_{\mathcal{V}_1} =\{i |d(\mathcal{V}_1,i)>L\}$; $j=2$. } \\
\textbf{1:} \textbf{while}{ $\mathcal{D}_{\mathcal{V}_1} \neq\emptyset$} \\
\textbf{2:} \quad $i=\min\{i | \alpha_i\in \mathcal{D}_{\mathcal{V}_1} \}$;\\
\textbf{3:} \quad $p_{\alpha_i} =q_{\alpha_j}$;\\
\textbf{4:} \quad $\mathcal{V}_1=\mathcal{V}_1\cup\{\alpha_i\}$; $\mathcal{D}_{\mathcal{V}_1} =\{i |d(\mathcal{V}_1,i)>L\}$; $j=j+1$; \\
\textbf{5:} \textbf{end while} \\
\textbf{6:} $n_1=|\mathcal{V}_1|$;\;\; $n_2=n-n_1$;\\
\textbf{7:} The remaining vertices in $\mathcal{V} \setminus\mathcal{V}_1$ are denoted by $\gamma_i, 1\leqslant i \leqslant n_2$, i.e. $\{\alpha_1,\alpha_2,\dots,\alpha_n\} \setminus\mathcal{V}_1=\{\gamma_1,\dots,\gamma_{n_2} \}$ in order;\\
\textbf{8:} $p_{\gamma_i} =q_{\alpha_{n_1+i}}, 1\leqslant i\leqslant n_2$;\\
\textbf{Output:}{ The reordered sampling probability distribution ${\bf p}^r=(p_1,p_2,\dots,p_n)$.} 
\label{code:recentEnd} 
\end{algorithm} 
The main idea of Algorithm \ref{reA} is that 
a vertex $u$ satisfying $\overline{\mathcal{N}}_u\cap \overline{\mathcal{N}}_v=\emptyset$ can be selected preferentially with large probability for all $v\in \mathcal{V}_1$. The preparation of this algorithm corresponds to computational cost $O(n\log{n})$, and the computational complexity in steps \textbf{1}-\textbf{5} is $O(n/\rho_L)$, where $\rho_L$ denotes the average cardinality of $\mathcal{B}(i,L)$ with $i\in\mathcal{V}$.
We remark that if the samples are selected in probability descending order without repetition, the output probability distribution in Algorithm \ref{reA} requires fewer samples than \eqref{optiP} to allow stable signal recovery by the definition of $\zeta_\Omega$ in \eqref{Gcoherence.def}.

\vspace{-1em}
\section{Numerical Experiments} \label{sec6} 
In this section, we demonstrate the effectiveness of the proposed optimal, estimated and reordered sampling probability from local weighted measurements by reconstructing the $k$-bandlimited graphs on a  random geometric graph and on a  monochrome image.

\vspace{-1em} 
\subsection{Local Random Sampling Scheme}
 Let  ${\bf p}_{\rm uni}$, ${\bf p}_{\rm opt}$ and ${\bf p}_{\rm est}$ denote  the uniform probability distribution, the optimal probability distribution in \eqref{optiP} and the estimated probability distribution in \eqref{estP}, respectively. Let ${\bf p}_{\rm opt}^r$ and ${\bf p}_{\rm est}^r$ further denote two reordered sampling probability distributions obtained from  ${\bf p}_{\rm opt}$ and ${\bf p}_{\rm est}$, respectively. In the following, we consider two weighted matrices ${\bf\Phi}_0={\bf I}$ and ${\bf\Phi}_1={\bf I}+{\bf L}$  in \eqref{phi2} for the experiments,  where the corresponding local weighted sampling matrices are  ${\bf\Psi}_0$ (the subset sampling (SS) method in \cite{gill1}) and ${\bf\Psi}_1$ (the local weighted sampling (LW) method), respectively. 
 
 \vspace{-1em} 
 \subsection{Sampling and Reconstruction on Random Geometric Graphs}
\begin{figure} [h]
\centering
\includegraphics[width=86mm, height=64mm]{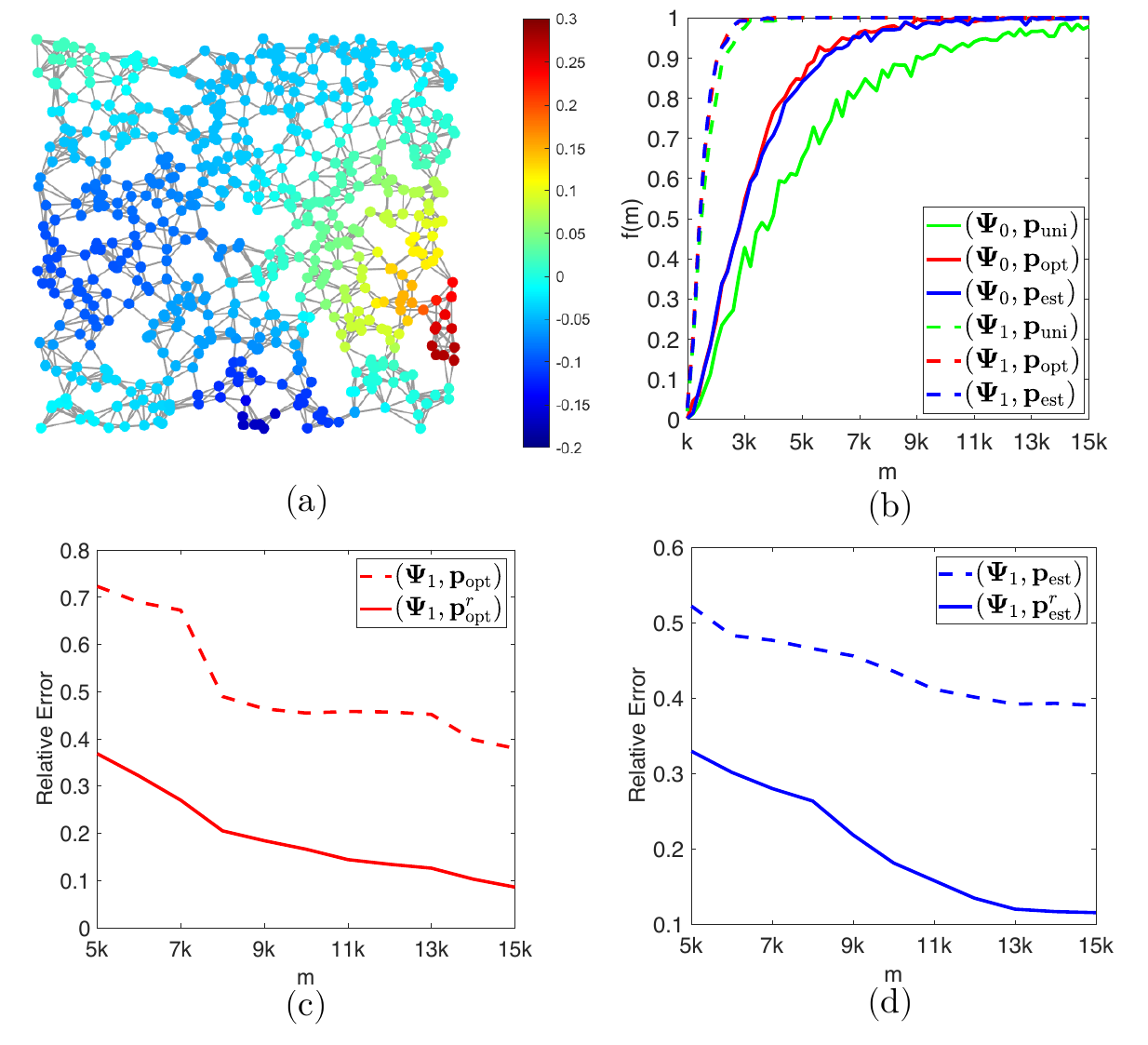}
\caption{\label{function} 
 (a) a $k$-bandlimited graph signal residing on a random sensor graph with 600 vertices and $k=10$;  (b) the probability functions $f(m)=\mathbb{P} (\underline{\delta}_k\leqslant 0.995)$; (c) the relative errors with  $({\bf \Psi}_1, {\bf p}_{\rm opt})$ (dashed line) and $({\bf \Psi}_1, {\bf p}^r_{\rm opt})$ (solid line);  
(d) the relative errors with $({\bf \Psi}_1, {\bf p}_{\rm est})$ (dashed line) and $({\bf \Psi}_1, {\bf p}^r_{\rm est})$ (solid line).
 } 
\end{figure}

 Let ${\mathcal G}_1$ be a random geometric graph \cite{gspbox} with $600$ vertices randomly deployed on $[0, 1]^2$. There exists an undirected edge between two vertices if they are within a fixed radius of each other, where the edge weight is assigned via a thresholded Gaussian kernel. Plotted in Figure \ref{function} (a) is a bandlimited signal ${\bf x} ={\bf U}_k\hat{\bf x}$ residing on $\mathcal{G}_1$ with bandwidth $k=10$, where the Fourier coefficients $\hat{\bf x}$ are randomly selected in $(-1, 1)$.
Based on \eqref{frame1} in Theorem \ref{main},  the lower bound of $\delta_k$ can be written as
 $ \underline{\delta}_k=\max\{1-\frac{\sigma_{\min} ({\bf P}_{\Omega}^{-\frac{1}{2}}{\bf\Psi}{\bf U}_k)}{mc_1}, \frac{\sigma_{\max} ({\bf P}_{\Omega}^{-\frac{1}{2}}{\bf\Psi}{\bf U}_k)}{mc_2} -1\}$,
 where $\Omega$ is the sampling set drawing by the probability distribution ${\bf p}_{\rm uni}$, ${\bf p}_{\rm opt}$, or ${\bf p}_{\rm est}$, and ${\bf\Psi}={\bf\Psi}_0$ or ${\bf\Psi}_1$.
  Similar to \cite{gill1}, 500 experiments are conducted for $k$-bandlimited signal to estimate the probability of $\underline{\delta}_k\leqslant 0.995$, i.e.,
  \begin{equation} \label{prob.m} 
f(m)=\mathbb{P} (\underline{\delta}_k\leqslant 0.995)=\frac{\sharp\{\underline{\delta}_k\leqslant 0.995\}}{\text{number of all experiments}},
\end{equation} 
 where $\sharp\{\underline{\delta}_k\leqslant 0.995\}$ denotes the number of experiments such that $\underline{\delta}_k\leqslant 0.995$. In particular, 
$f(m)=1$ implies that \eqref{frame1}  holds.  
From Plot (b) in Figure \ref{function}, we observe that the function $f(m)$ with ${\bf p}_{\rm opt}$ or ${\bf p}_{\rm est}$ needs  fewer measurements to reach $1$ than that with ${\bf p}_{\rm uni}$ under the same sampling method (${\bf \Psi}_0$ or ${\bf \Psi}_1$).  Also, if the sampled nodes are drawn by the same probability distribution, the measurements number taken with ${\bf \Psi}_1$ (LW) is smaller than that with ${\bf \Psi}_0$ (SS) to reach $f(m)=1$. 

We now demonstrate the effectiveness of the  proposed reordered sampling probability distribution in Algorithm \ref{reA} by the relative error ${\|{\bm x}^{*}-{\bm x}\|_2}/{\|{\bm x}\|_2}$, where 
\begin{equation}\label{solutionx}
{\bm x}^{*}=({\bf \Psi}^T{\bf P}_\Omega^{-1}{\bf \Psi}+{\bf L})^{-1}({\bf \Psi}^T{\bf P}_\Omega^{-1}{\bf y})
\end{equation}
 is the reconstruction obtained through solving $\min_{{\bf x}\in \mathbb{R}^n} \frac{1}{2}\|{\bf P}^{-1/2}_\Omega({\bf \Psi}{\bf x}-{\bf y})\|_2^2+\frac{1}{2}{\bf x}^T{\bf L}{\bf x},$
and ${\bf y}\in \mathbb{R}^m$ is in \eqref{sampleddata}, cf. \cite{gill1}. From plots (c) and (d) in Figure \ref{function}, we have that with the local weighted sampling matrix ${\bf \Psi}_1$, the  samples drawn by the  reordered probability distributions ${\bf p}^r_{\rm opt}$ and ${\bf p}^r_{\rm est}$ provide a better approximation to the original $k$-bandlimited graph signal than that obtained from samples drawn by the probability distributions ${\bf p}_{\rm opt}$ and ${\bf p}_{\rm est}$.  
\vspace{-1.2em} 
\vspace{-2em} 
\subsection{Random Sampling on Monochrome Image}\label{realdata} 
\begin{figure}[h]
\centering
\includegraphics[width=86mm, height=64mm]{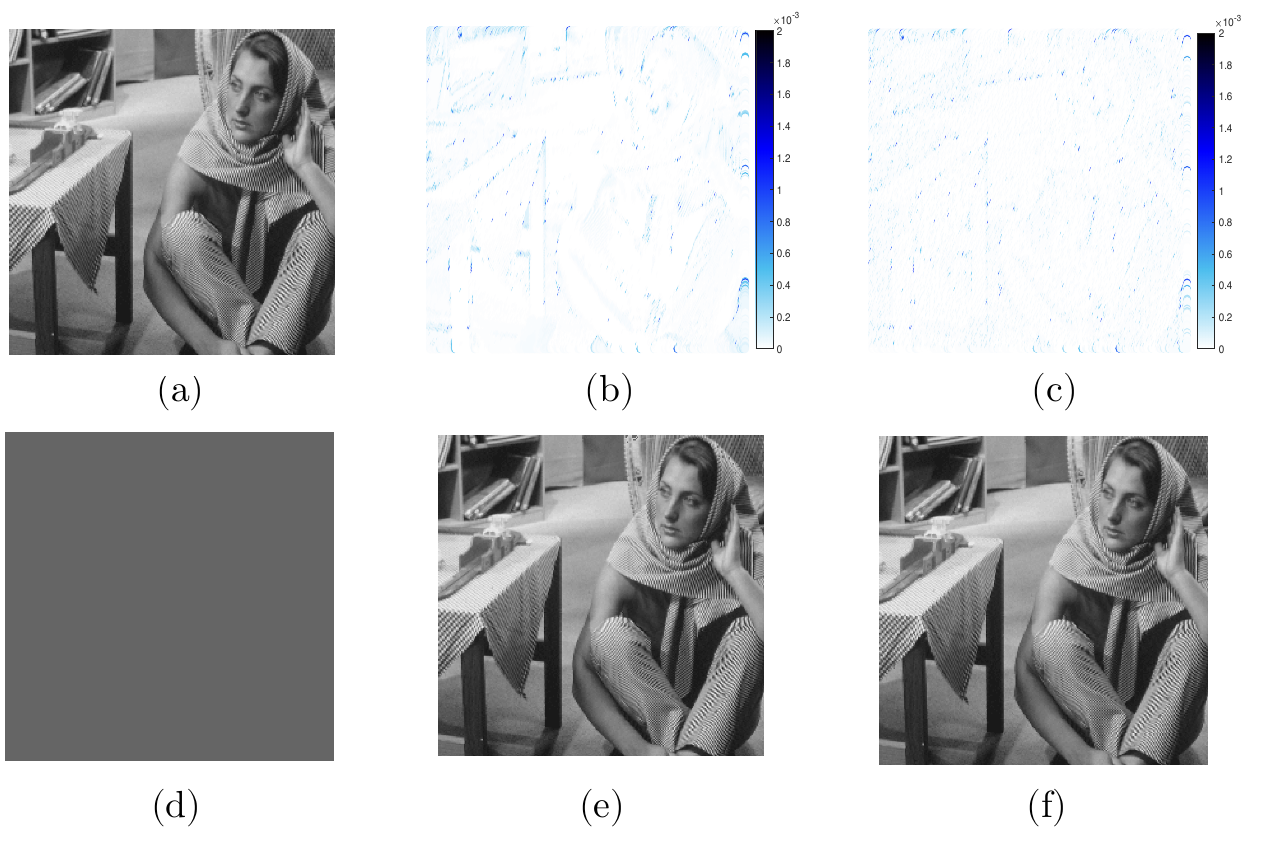}\\
\caption{\label{pepper.pd} Plotted on the top from the left to the right are the original image, the  estimated sampling distribution ${\bf p}_{\rm est}$, and the reordered estimated optimal sampling distribution ${\bf p}^r_{\rm est}$, resp. 
Plotted on the bottom from the left to the right are the reconstructions from samples drawn  by $({\bf \Psi}_1,{\bf p}_{\rm uni})$ (the output SNR=0 dB),   $({\bf \Psi}_1,{\bf p}_{\rm est})$ (the output SNR=$14.33$ dB), and   $({\bf \Psi}_1,{\bf p}^r_{\rm est})$ (the output SNR=$17.32$ dB), where $k=236$ and $m=3540$. } 
\end{figure}

In this subsection, the experiments are conducted based on the Barbara picture in \cite{barbara.database}, see Figure \ref{pepper.pd} (a). The monochrome image contains ${256\times 256}$ pixels, and can be represented by a matrix ${\bm X}\in \mathbb{R}^{256\times 256}$. An undirected graph model is built with $n=65536$ vertices represented all pixels, and the edges are constructed by the 10 nearest neighboring algorithm on the coordinates of picture pixels. 
Define the signal-to-noise ratio 
$\text{SNR}=-10\log_{10}\frac{\|{\bm X}^*-{\bm X}\|_F}{\|{\bm X}\|_F}$, where ${\bm X}^*$ is the reconstruction obtained by \eqref{solutionx}. 

\begin{table}[h]
  \centering
  \captionsetup{justification=centering}
  \caption{\label{table}Some signal energy ratios concentrated on the first $k$ Fourier modes $Eg_k=\frac{\|{\bf U}_{k}{\bf U}_{k}^T{\bm X}\|_F}{\|{\bm X}\|_F}$ with $k=\lceil \frac{i*n}{10000}\rceil$.}
  \setlength{\tabcolsep}{1.95mm}{
  \begin{tabular}{|c|ccccccc|}
     \hline
     $i$ &16 & 21 &26 &31 &36&41 & 46\\
    \hline
     $k$ &105 & 138 & 171 &204&236&269&302\\
    \hline
    $Eg_k$& 0.52 & 0.66 & 0.76 &0.83&0.92&0.94&0.99\\
    \hline
  \end{tabular}
  }
\end{table}

Based on Table \ref{table},  we set the bandwidth $k=236$ for the graph signal ${\bm X}$. In the following, we reconstruct the Barbara image from $m=15k=3540$ measurements drawn by the uniform distribution ${\bf p}_{\rm uni}$, the estimated distribution ${\bf p}_{\rm est}$ in \eqref{estP}, and the reordered estimated distribution ${\bf p}^r_{\rm est}$ in Algorithm \ref{reA}, where the local weighted sampling matrix is ${\bf \Psi}_1$ (LW), see Figure \ref{pepper.pd}. We observe that the reordered random sampling scheme $({\bf \Psi}_{1}, {\bf P}^r_{\rm est})$ yields a better reconstruction than the random sampling scheme $({\bf \Psi}_{1}, {\bf P}_{\rm est})$, where the output SNRs are  $17.32$ dB and $14.33$ dB, respectively.  However, the random sampling scheme $({\bf \Psi}_{1}, {\bf P}_{\rm uni})$ fails to lead a reconstruction. This shows that reordered estimated sampling probability distribution  improves the performance of the estimated sampling probability distribution, and both of them outperform the uniform sampling distribution. 
 
We further reconstruct the Barbara picture dataset from $m=15k=3540$ measurements drawn by the sampling probability distributions ${\bf p}_{\rm uni}, {\bf p}_{\rm est}, {\bf p}^r_{\rm est}$ with the subset sampling matrix ${\bf \Psi}_0$ (SS). 
We observed that the $({\bf \Psi}_0, {\bf p}_{\rm uni})$ fails to provide an approximation, and the SNRs of $({\bf \Psi}_0,{\bf p}_{\rm est})$ and $({\bf \Psi}_0,{\bf p}^r_{\rm est})$ are  $13.60$ dB and $13.67$ dB correspondingly.  This implies that both the reordered estimated sampling and the estimated sampling distributions perform better than the uniform sampling distribution in the reconstruction. However, the reordered estimated sampling sampling can not work for SS due to the positive integer $L=0$ in Algorithm \ref{reA}.

\vspace{-0.5em} 
\vspace{-1em}

\vspace{-1em} 
\section{Conclusion} \label{sec7} 
In this letter, we proposed a new random sampling scheme for $k$-bandlimited graph signals. It was shown that the $k$-bandlimited graph signal can be recovered from local measurements with high probability if the number of measurements is $O(k\log k)$. Based on this, the probability distribution of optimal sampling  and estimated sampling were obtained. 
A new sampling algorithm was further proposed to reduce the redundancy of signal values. 
Simulations verified that the proposed methods yield a better performance than the existing ones in the stable recovery.

\begin{appendix}[Supplementary Material]
\subsection{Proof of Theorem \uppercase\expandafter{\romannumeral 2}.1 } \label{sec1}
\begin{proof}
For any ${\bf x} \in \text{span}({\bf U}_k)$, i.e., ${\bf x} ={\bf U}_k\hat{\bf x}$ with $\hat{\bf x} \in {\mathbb R}^k$, we have $$\|{\bf P}_\Omega^{-\frac{1}{2}}{\bf\Psi}{\bf x} \|_2^2=\hat{\bf x}^T({\bf P}_\Omega^{-\frac{1}{2}}{\bf\Psi}{\bf U}_k)^{T}({\bf P}_\Omega^{-\frac{1}{2}}{\bf\Psi }{\bf U}_k)\hat{\bf x}.$$
 Let us define
 \begin{equation} 
\begin{aligned} 
{\bf X} &=\frac{1}{m} ({\bf P}_\Omega^{-\frac{1}{2}}{\bf\Psi}{\bf U}_k)^{T}({\bf P}_\Omega^{-\frac{1}{2}}{\bf\Psi }{\bf U}_k)\\
&=\frac{1}{m}{\bf U}_k^{T}\sum_{i=1}^m{\frac{{\pmb \varphi}_{\omega_i}{\pmb \varphi}_{\omega_i}^{T}}{p_{\omega_i}} }{\bf U}_k\\
&:=\sum_{i=1}^m{\bf X}_{\omega_i},
\end{aligned} 
\end{equation} 
where
\begin{equation} 
\begin{aligned} 
{\bf X}_{\omega_i} ={\bf U}_k^{T}{\frac{{\pmb \varphi}_{\omega_i}{\pmb \varphi}_{\omega_i}^{T}}{mp_{\omega_i}} }{\bf U}_k=\frac{1}{m}\Big({\frac{{\bf U}_k^{T}{\pmb \varphi}_{\omega_i}}{p^{-1/2}_{\omega_i}} }\Big)\Big({\frac{{\bf U}_k^{T}{\pmb \varphi}_{\omega_i}}{p^{-1/2}_{\omega_i}} }\Big)^{T}. 
\end{aligned} 
\end{equation} 
Then, since all ${\bf X}_{\omega_i}$ for $1\leqslant i\leqslant m$ are self-adjoint, positive semidefinite $k\times k$ matrices, we have
\begin{equation} \nonumber
\begin{aligned} 
\lambda_{\max} ({\bf X}_{\omega_i})&=\|{\bf X}_{\omega_i} \|_2\\
&\leqslant \max\limits_{{\omega_i} \in \Omega} \|{\bf X}_{\omega_i} \|_2\\
&=\frac{1}{m} \max\limits_{{\omega_i}\in\Omega} \{\|{\frac{{\bf U}_k^{T}{\pmb \varphi}_{\omega_i}}{p^{-1/2}_{\omega_i}}}\|^2_2\}, .
\end{aligned} 
\end{equation} 
Taking $\zeta_\Omega=\max_{{\omega_i}\in\Omega} \{{\frac{\|{\bf U}_k^{T}{\pmb \varphi}_{\omega_i}\|^2_2}{p_{\omega_i}}}\}$, 
the inequality mentioned above can be formulated as
\begin{equation} \label{pr1.1} 
\begin{aligned} 
\lambda_{\max} ({\bf X}_{\omega_i})\leqslant\frac{\zeta_\Omega}{m}.
\end{aligned} 
\end{equation} 
Since each node $\omega_i$ in the sampling set $\Omega$ is randomly and independently selected from $\{1,2,\dots,n\}$ with probability distribution $\bf p$,  we have
\begin{equation} \nonumber
\begin{aligned} 
\mathbb{E} ({\bf X}_{\omega_i})&=\frac{1}{m}{\bf U}_k^{T} \mathbb{E} ({\frac{{\pmb \varphi}_{\omega_i}{\pmb \varphi}_{\omega_i}^{T}}{p_{\omega_i}} }){\bf U}_k\\
&=\frac{1}{m}{\bf U}_k^{T}\Big(\sum_{i=1}^n p_i{\frac{{\pmb \varphi}_i{\pmb \varphi}_i^{T}}{p_{i}} } \Big){\bf U}_k\\
&=\frac{1}{m}{\bf U}_k^{T} {\bf\Phi}^{T}{\bf\Phi}{\bf U}_k\\
&=\frac{1}{m}{\bf U}_k^{T} {\bf U} (g({\bf\Lambda}))^2{\bf U}^{T}{\bf U}_k\\
&=\frac{1}{m}\text{diag} (g(\lambda_1)^2, g(\lambda_2)^2,\dots, g(\lambda_k)^2)
\end{aligned} 
\end{equation} 
for all ${\bf X}_{\omega_i}, 1\leqslant i \leqslant m$. 
This implies that
\begin{equation} \label{pr1.2} 
\begin{aligned} 
\mu_{\min}=\lambda_{\min} (\sum_i\mathbb{E} ({\bf X}_{\omega_i}))=\min_{1\leqslant i\leqslant k} \{(g(\lambda_i))^2\},
\end{aligned} 
\end{equation} 
and
\begin{equation} \label{pr1.3} 
\begin{aligned} 
\mu_{\max} =\lambda_{\max} (\sum_i\mathbb{E} ({\bf X}_{\omega_i}))=\max_{1\leqslant i\leqslant k} \{(g(\lambda_i))^2\}.
\end{aligned} 
\end{equation} 
Substitute \eqref{pr1.1}, \eqref{pr1.2} and \eqref{pr1.3} into [37, Theorem 1.1], for any $\delta\in(0,1)$, we obtain 
\begin{align*} 
\mathbb{P} \big[\lambda_{\min} ({\bf X})\leqslant(1-\delta)\mu_{\min} \big]&\leqslant k[\frac{e^{-\delta}}{(1-\delta)^{1-\delta}}]^{\frac{\mu_{\min} m}{\zeta_\Omega}}\\
& \leqslant k \exp(-\frac{\delta^2\mu_{\min} m}{3\zeta_\Omega})
\end{align*}
and
\begin{align*} 
\mathbb{P} \big[\lambda_{\max} ({\bf X})\geqslant(1+\delta)\mu_{\max} \big]&\leqslant k[\frac{e^{\delta}}{(1+\delta)^{1+\delta}}]^{\frac{\mu_{\max} m}{\zeta_\Omega}} \\
&\leqslant k \exp(-\frac{\delta^2\mu_{\min} m}{3\zeta_\Omega}).
\end{align*} 
Obviously, $k \exp(-\frac{\delta^2\mu_{\min} m}{3\zeta_\Omega})\leqslant\frac{\varepsilon}{2}$ always holds if 
\begin{align*} 
m\geqslant \frac{3\zeta_\Omega }{\mu_{\min} \delta^2} \log\frac{2k}{\varepsilon}.
\end{align*} 
Thus, with probability at most $\varepsilon$,
\[\lambda_{\min} ({\bf X})\leqslant(1-\delta)\mu_{\min} \;\;\text{or} \;\;\lambda_{\max} ({\bf X})\geqslant(1+\delta)\mu_{\max},\]
which states that with probability at least $1-\varepsilon$,
\[\lambda_{\min} ({\bf X})\geqslant(1-\delta)\mu_{\min} \text{ and } 
\lambda_{\max} ({\bf X})\leqslant(1+\delta)\mu_{\max}.\]
 Combined Rayleigh quotient  with $\|\hat{\bf x} \|_2^2=\|{\bf U}_k\hat{\bf x} \|_2^2=\|{\bf x} \|_2^2$, it can be seen that the following inequality
 \begin{align*} 
 (1-\delta)\mu_{\min} \|\hat{\bf x} \|_2^2&\leqslant\langle \hat{\bf x}, \frac{1}{m} ({\bf P}_\Omega^{-\frac{1}{2}}{\bf\Psi}{\bf U}_k)^{T}({\bf P}_\Omega^{-\frac{1}{2}}{\bf\Psi}{\bf U}_k)\hat{\bf x} \rangle\\
& \leqslant(1+\delta)\mu_{\max} \|\hat{\bf x} \|_2^2
\end{align*} 
holds with $m\geqslant \frac{3\zeta_\Omega}{c_1\delta^2} \log\frac{2k}{\varepsilon}$. 
Taking 
$c_1=\mu_{\min}$ in \eqref{pr1.2} 
and $c_2=\mu_{\max}$ in \eqref{pr1.3}, 
the inequality above is equivalent to
\begin{align*} 
 (1-\delta)c_1\|{\bf x} \|_2^2 &\leqslant \|{\bf P}_\Omega^{-\frac{1}{2}}{\bf\Psi}{\bf x} \|_2^2\leqslant(1+\delta)c_2 \|{\bf x} \|_2^2,
\end{align*} 
for all ${\bf x}\in\text{span}({\bf U}_k)$.
Therefore, the proof is completed.
\end{proof}

\subsection{Signal Energy Ratio in Numerical Experiments}
Let ${\bf U}_{k}{\bf U}_{k}^T{\bm X}$ be the orthogonal projection of ${\bm X}$ onto $\text{span}({\bm U}_k)$. Table \ref{table} shows some signal energy ratios ($Eg_k=\frac{\|{\bf U}_{k}{\bf U}_{k}^T{\bm X}\|_F}{\|{\bm X}\|_F}$) concentrated on the first $k$ Fourier modes with different $k=\lceil \frac{i*n}{10000}\rceil$ in Barbara image. 
It should be noticed from Table \ref{table} that the value of $k$ can be within a large range from $236$ for the small variations in signal energy ratio $Eg_k$ to be small, so the bandwidth $k=236$ can be taken in our experiments.
\begin{table}[h]
  \centering
  \captionsetup{justification=centering}
  \caption{\label{table}The signal energy ratio($Eg_k$) concentrated on the first $k$ Fourier modes where $Eg_k=\frac{\|{\bf U}_{k}{\bf U}_{k}^T{\bm X}\|_F}{\|{\bm X}\|_F}$ with $k=\lceil \frac{i*n}{10000}\rceil$.}
  \begin{tabular}{|c|ccccccc|}
     \hline
     $i$ &1 & 6&11 &16&21 & 26&31 \\
          &36&41 & 46&51&56&61&\dots\\
    \hline
     $k$ &7 & 40&73 &105&138&171&204\\
           &236&269&302&335&368&400&\dots\\
    \hline
    $Eg_k$& 0.20 & 0.28&0.36&0.52&0.66&0.76&0.83\\
    &0.92&0.94&0.99&1&1&1&\dots\\
    \hline
  \end{tabular}
\end{table}

\end{appendix}

\end{document}